\documentclass[preprint,aps,nofootinbib,floatfix,superscriptaddress,longbibliography]{revtex4-1}
\usepackage{amsmath,amsfonts,amssymb,amsthm,
                    mathtools, 
					  esint, 
				    hyperref, 
			      mathrsfs, 
		        graphicx} 

\usepackage[nodisplayskipstretch]{setspace}
\usepackage{orcidlink}
\usepackage[utf8]{inputenc}
\usepackage[T1]{fontenc}
\usepackage{graphicx}
\usepackage[version=4]{mhchem}
\usepackage{stmaryrd}
\usepackage{amsbsy}
\usepackage{latexsym}
\usepackage{comment}
\usepackage{natbib}
\usepackage{bm}
\usepackage{subfigure} 
\usepackage{color}
\usepackage{wasysym}
\usepackage{mathbbol}
\usepackage{bigints}
\allowdisplaybreaks
\usepackage[normalem]{ulem}
\usepackage[dvipsnames]{xcolor}
\usepackage{multirow}
\usepackage{physics}
\usepackage[export]{adjustbox}
\graphicspath{ {./images/} }

\begin{document}
\title{Cosmological Evolution of Primordial Black Holes: UV/IR Decoupling and the KM3NeT 220 PeV Neutrino Anomaly} 
\author{Debdatta Dey}
\thanks{Equal contribution}
\email{deb.datta.dey2002@gmail.com}
\affiliation{Department of Physics, Birla Institute of Technology and Science,
Pilani-Hyderabad Campus Hyderabad 500078, India}
\affiliation{Department of Physics,  Indian Institute of Technology Bombay, Mumbai 400076, India}
\author{Tausif Parvez}
\thanks{Equal contribution}
\email{214120002@iitb.ac.in}
\affiliation{Department of Physics,  Indian Institute of Technology Bombay, Mumbai 400076, India}
\author{S. Shankaranarayanan \orcidlink{0000-0002-7666-4116}} 
\email{shanki@iitb.ac.in}
\affiliation{Department of Physics,  Indian Institute of Technology Bombay, Mumbai 400076, India}
\begin{abstract}
The recent observation of a 220 PeV neutrino event (KM3-230213A) by the KM3NeT observatory presents a formidable challenge to standard astrophysical source models. We investigate the hypothesis that this ultra-high-energy signature originates from the terminal evaporation burst of a Primordial Black Hole (PBH). Since PBH evolution spans cosmic history, static vacuum approximations fail to capture early-universe dynamics. Embedding the PBH in a cosmological background via the \emph{McVittie spacetime}, we demonstrate that early-universe cosmological accretion and expansion-suppressed Hawking emission shift the required initial mass window for a terminal burst occurring today. We show that although the early universe environment dictates the black hole's overall lifespan, its final explosion today ($z \approx 0$) is governed by standard Schwarzschild thermodynamics. This mechanism naturally produces the intense 220 PeV local flux while suppressing early emissions, thereby satisfying diffuse isotropic background limits. Consequently, this dynamical framework alters the mapping between current ultra-high-energy neutrino observables and the primordial curvature perturbations that seeded them.
\end{abstract}

\maketitle

\section{Introduction}
\label{sec:introduction}

Primordial Black Holes (PBHs) constitute a unique probe of early-universe cosmology, general relativity, and quantum field theory~\cite{Zeldovich:1967lct,Hawking:1971ei,Carr:1974nx,Carr:1975qj}. Formed from the collapse of extreme density fluctuations in the radiation-dominated epoch, their subsequent evolution spans vast cosmological timescales and leading to potential observational consequences. While macroscopic PBHs are frequently studied as dark matter candidates or seeds for large-scale structure, the evolution and final stages of lighter PBHs are distinctly governed by semi-classical gravity~\cite{Carr:2020xqk,Shankaranarayanan:2026hnn}. For these lower-mass remnants, quantum effects at the event horizon lead to the continuous emission of a nearly thermal spectrum of Hawking 
particles~\cite{Hawking:1975vcx,Page:1976df,Page:1977um,Damour:1976jd,Shankaranarayanan:2000qv}. As the evaporation proceeds and the BH mass shrinks, the emission rate exponentially accelerates, providing a theoretical framework for investigating potential localized sources of high-energy astrophysical particles~\cite{Halzen:1991uw,Bambeck:2005bz,Carr:2009jm,Lunardini:2019zob,Auffinger:2020afu,Auffinger:2022khh,Bernal:2022swt,Liu:2023cqs,Zantedeschi:2024ram}.

While Hawking evaporation can theoretically produce the entire spectrum of Standard Model particles, the emission probability is exponentially suppressed for particles heavier than the BH's instantaneous surface temperature, massless particles and those with tiny masses --- \emph{like Neutrinos} --- are the predominantly produced species over majority of the BH's lifetime~\cite{Page:1976df}. Further, while ultra-high-energy (UHE) photons are also emitted during the end stages of evaporation, they interact strongly with the interstellar medium and the cosmic microwave background, rapidly cascading into secondary cosmic rays. In contrast, neutrinos interact solely via the weak force and have negligible interaction cross-sections over cosmological distances and dense local environments. This makes them the most robust direct messengers for isolating UHE PBH signatures on Earth.

The operation of large-volume neutrino telescopes, including IceCube and KM3NeT~\cite{Halzen:2002pg,IceCube-Gen2:2020qha,Katz:2006wv,KM3NeT:2018wnd}, has extended the observational reach to UHE neutrinos~\cite{IceCube:2013cdw,IceCube:2020wum}. A notable event in this regime is KM3-230213A, a track-like signature with an estimated energy of approximately $220\text{ PeV}$~\cite{KM3NeT:2025npi}. This event is distinguished from the diffuse astrophysical neutrino background by its apparent point-source origin. If the event is interpreted as the final emission from a nearby evaporating PBH, its properties strongly disagree with the predictions of the static Schwarzschild metric~\cite{Airoldi:2025opo,Baker:2025cff,Brdar:2025azm}.

In a static metric, the PBH mass loss is governed by the Page approximation for Hawking radiation~\cite{Page:1976df}:
\begin{equation}
\label{eq:MassEvol-Sch}
    \left( \frac{dM}{dt} \right)_{\text{Sch}} = - \frac{\alpha_{\text{evap}}}{M^2}
\end{equation}
where $\alpha_{\text{evap}}$ is a mass-dependent coefficient accounting for the degrees of freedom of emitted particles. 
Integrating the Schwarzschild mass loss rate from the time of formation $t_i$ to the present day $t_0$ naively requires an initial mass of $M_{\text{in,Sch}} \approx (3 \alpha_{\text{evap}} t_0)^{1/3} \approx 5 \times 10^{14}\text{ g}$ to yield a terminal burst today. A PBH must attain a critically low mass of $M \approx 10^{4}\text{ g}$ to reach an instantaneous Hawking temperature corresponding to energy of the order 200 $\text{ PeV}$~\cite{MacGibbon:1990zk}. Since, the static evaporation rate diverges as the mass approaches zero, this high-temperature phase is transient, lasting less than $\Delta t \sim 10^{-10}\text{ seconds}$~\cite{MacGibbon:1991tj,Carr:1998fw}.

For a single PBH at luminosity distance $d_L(z_s)$, the observed point-source photon flux is determined by the emission integrated over this brief interval:
\begin{equation}
F(E_0) = \frac{1}{4\pi d_L^2(z_s)} \int_{t_{burst}-\Delta t}^{t_{burst}} \frac{d^2N}{dE dt} \bigg\vert{}_{E = E_0(1+z_s)} dt \, .
\end{equation}
Particles emitted at early times undergo cosmological redshift ($E = E_0(1+z)$). Given that $\Delta t < 10^{-10}\text{ s}$, the predicted flux is severely suppressed, requiring a highly unnatural local overdensity of PBHs to make the detection of a $220\text{ PeV}$ neutrino statistically probable.
Further, this limitation impacts predictions for the diffuse UHE neutrino flux. For a cosmological population of PBHs with comoving number density $n_{PBH}(z)$, the expected diffuse differential flux $\Phi(E_0)$ integrates the emission spectrum over cosmic history~\cite{Lunardini:2019zob}:
$$\Phi(E_0) = \frac{c}{4\pi} \int_{0}^{z_{in}} \frac{dz}{H(z)(1+z)} n_{PBH}(z) \frac{d^2N}{dE dt} \bigg\vert{}_{E = E_0(1+z)}$$
where $H(z)$ is the Hubble parameter. In the standard scenario, the energy-scaled flux $E_0^2 \Phi(E_0)$ decreases at UHEs and fails to account for PeV-scale signatures without violating established background limits~\cite{Wu:2024uxa,Mukhopadhyay:2026lmz}.

This discrepancy reveals a limitation in modeling of PBH as isolated objects in an asymptotically flat space-time. They reside in an expanding Friedmann-Lemaitre-Robertson-Walker (FLRW) universe~\cite{Shankaranarayanan:2026hnn}. 
%
To accurately map the currently observed terminal burst to the primordial curvature perturbations that generated the PBH, one must track the exact mass evolution over $13.8$ billion years. However, a static Schwarzschild analysis discussed above is  insufficient, as it decouples the local horizon dynamics from the expanding FLRW background. 

In this work, by modeling a dynamical background metric via the McVittie solution~\cite{McVittie:1933zz}, we embed the PBH within the cosmic fluid. This approach allows us to capture two dominant contributions in the early universe: accretion of the radiation bath~\footnote{While the strict McVittie metric does not support radial fluid accretion, operating in a quasi-stationary adiabatic limit effectively treats the background as a piecewise sequence of McVittie slices. This provides a rigorous approximation of the accretion dynamics inherent to a generalized-Vaidya spacetime, ensuring a consistent phase-space integration over the entire cosmic history.} and the modification of the apparent horizon surface gravity. By evaluating the semi-classical tunneling probability across the apparent horizon, we demonstrate that the effective Hawking temperature ($T_H$) acquires a dynamic correction explicitly dependent on the cosmological expansion rate $H(t)$. Consequently, the PBH evolution is governed by a competition between this dynamically modified Hawking evaporation and cosmological fluid accretion.

Our analysis reveals that these dynamic corrections, while negligible in the late universe ($z \to 0$), have appreciable contribution in the early universe. The high energy density in the early universe leads to a rapid accretion phase, suppresses early-time UHE evaporation, inherently preserving the diffuse isotropic neutrino limits. We thus establish that a proper ultraviolet (UV) and infrared (IR) scale decoupling alters the mapping between observed local point-source events --- such as KM3-230213A --- and the requisite early-universe curvature perturbations, offering a mathematically consistent framework for cosmological PBH phenomenology.

\section{Cosmological Black Hole spacetime}
\label{sec:geometry}

To model the evaporation of a PBH in an expanding universe, we employ the McVittie metric~\cite{McVittie:1933zz,Kaloper:2010ec,Gaur:2022hap}, which provides an exact solution to Einstein's field equations for a spherically symmetric mass $M$ embedded in a spatially flat FLRW background (corresponding to time-dependent homogeneous density and pressure). Specifically, we employ the adiabatic approximation, treating the mass $M(t)$ as evolving slowly compared to the light-crossing time of the horizon ($\dot{M} \ll 1$). The spacetime is thus assumed to relax to the instantaneous McVittie form at each moment.

While traditionally expressed in isotropic coordinates, analyzing particle emission across dynamical horizons requires a coordinate system that is regular at the apparent horizon. To go about that, we transform from the isotropic radial coordinate $R$ to the areal radius $r = a(t)R \left( 1 + \frac{M}{2a(t)R} \right)^2$, where $M$ is the BH mass and $a(t)$ is the scale factor. As discussed in Appendix \eqref{app:mcvittie_details}, this introduces a non-diagonal cross-term, casting the McVittie metric into a generalized Painlev\'e-Gullstrand-like form~\cite{Gaur:2022hap}: (We use geometric units where $G = c = 1$.) 
\begin{align}
\label{eq:McVittie_PG}
ds^2 &= - \left(f(r) - H^2(t) r^2 \right) dt^2 - {2 H(t) r} \, dt \, dr/{\sqrt{f(r)}} \nonumber \\  
&+ {dr^2}/{f(r)} + r^2 d\Omega^2
\end{align}
where $H(t) = \dot{a}(t)/a(t)$ is the Hubble parameter, 
$f(r) \equiv 1 - {2M}/{r}$, and $d\Omega^2 = d\theta^2 + \sin^2\theta d\phi^2$. This metric isolates the two-dimensional normal space $ds^2 = h_{ab} dx^a dx^b + r^2 d\Omega^2$ in coordinates $x^a \in \{t, r \}$.  The off-diagonal $dt \, dr$ term encapsulates the dynamical coupling between the gravitational pull of the BH and the cosmic expansion.
\begin{figure}
\centering
\includegraphics[width=0.80\textwidth]{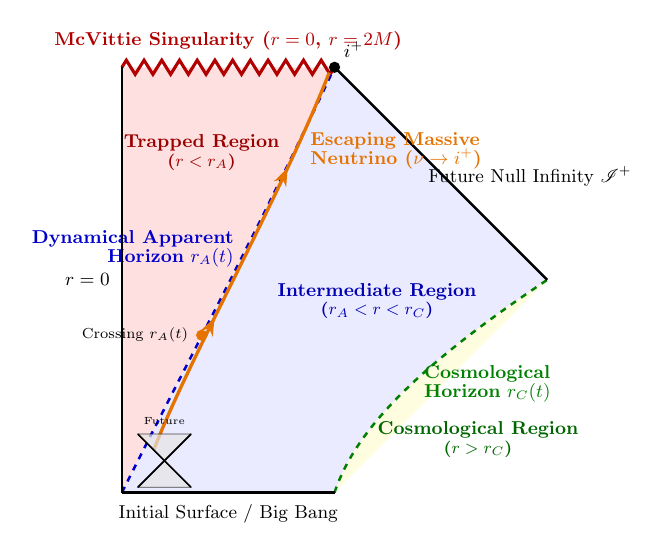}
\caption{\label{fig:Penrose} 
Penrose-Carter diagram of the McVittie spacetime. $H(t)$ necessitates a spacelike big bang singularity and time-dependent horizons. While the worldlines of UHE neutrino terminate at future timelike infinity ($i^+$), their extreme Lorentz factors ($\gamma \gg 1$) cause their trajectories to track arbitrarily close to the null boundary ($\mathscr{I}^+$) on cosmological scales.
}     
\end{figure}

To determine the causal boundaries, we evaluate the condition for the apparent horizon, defined by the vanishing of the radial flux~\cite{Wald:1984rg,Faraoni:2015ula}: 
\begin{equation}
    \label{eq:Apparent_Horizon}
    f(r_A) =  H^2(t) r_A^2 \, .
\end{equation}
For $3\sqrt{3}MH(t) < 1$, this cubic equation admits two real, positive roots. As seen from Fig.~\eqref{fig:Penrose}, these represent the BH apparent horizon and the cosmological horizon.

Since, the apparent horizon ($r_A$) evolves dynamically, the associated surface gravity cannot be calculated via the standard static Killing vector formalism. Instead, we utilize Hayward's trapping formalism~\cite{Hayward:1993wb,Hayward:1997jp} based on the geometric Kodama vector~\cite{Kodama:1979vn,Abreu:2010ru,Racz:2005pm} (detailed in Appendix \ref{app:mcvittie_details}). The invariant dynamical surface gravity evaluated at the BH apparent horizon is:
\begin{equation}
\label{eq:Surface_Gravity}
\kappa_A = \frac{M}{r_A^2} - r_A H^2(t) - \frac{\dot{H}(t)}{2H(t)}~.
\end{equation}
The first term is the standard Schwarzschild contribution, the second represents the local suppression of surface gravity due to the expansion, and the third $\dot{H}$ term contains the changing Hubble rate. As we will see, it is this time-dependent geometry that modifies the emission rate in the early universe.

\section{Fermionic Emission in a Dynamical Background}
\label{sec:fermionic_emission}

To analyze the emission of Fermions from the apparent horizon, we consider the covariant Dirac equation 
\begin{equation}
    i\gamma^\mu\mathcal{D}_\mu\psi + {m} \psi/{\hbar} = 0~,
\end{equation}
in the above line-element \eqref{eq:McVittie_PG}. Here, $\gamma^\mu = e^\mu_a \gamma^a$ are the curved spacetime gamma matrices, $\mathcal{D}_\mu = \partial_\mu - \frac{1}{4}\omega_{\mu ab}\gamma^a\gamma^b$ is the spinor covariant derivative,  $e^\mu_a$ are the tetrad fields, and $m$ is the Fermion (neutrino) mass. As detailed in Appendix \ref{app:dirac_separation}, the background spherical symmetry allows us to decouple the angular dependence using spin-weighted spherical harmonics, reducing the system to a set of coupled partial differential equations (see Appendix \ref{app:field_simplification}).

To extract the semi-classical particle emission spectrum, we employ the WKB approximation~\cite{Damour:1976jd,Zhao:1994sc,Srinivasan:1998ty,Shankaranarayanan:2000gb,Shankaranarayanan:2000qv,Shankaranarayanan:2003ya}. We apply the adiabatic ansatz for the up-spin (and similarly down-spin) spinor field:
\begin{equation}
    \label{spinor_ansatz}
    \psi_{\uparrow}(t,r,\theta,\phi) = \exp\left[\frac{i}{\hbar} I_{\uparrow}(t,r,\theta,\phi)\right] 
    \begin{bmatrix}
        A(t,r,\theta,\phi) \\ 0 \\ B(t,r,\theta,\phi) \\ 0
    \end{bmatrix}~.
\end{equation}
In a dynamical spacetime lacking a globally timelike Killing vector, the preferred notion of energy for the emitted particle is defined via the Kodama vector field $K^\mu$~\cite{Kodama:1979vn,Shankaranarayanan:2026hnn}. The invariant Kodama energy is $\omega(t,r) = -K^\mu \partial_\mu I_{\uparrow} = -\partial_t I_{\uparrow}$~\cite{Xavier:2021chn}. 

We focus on radial emission ($J=0$) in the massless limit ($m=0$), which dominates the Hawking radiation spectrum near the horizon~\cite{Page:1976df,Page:1977um}. Substituting Eq.~\eqref{spinor_ansatz} into the Dirac equation and expanding to lowest order in $\hbar$ yields a system of algebraic equations for the amplitudes $A$ and $B$. Enforcing non-trivial solutions leads to two distinct branches (corresponding to outgoing and ingoing modes~\cite{Shankaranarayanan:2003ya}) for the effective radial momentum, $k_\pm(r, t) \equiv \partial_r I_{\uparrow}^{\pm}(r, t)$:
\begin{align}
\label{eq:effective_momentum_plus}
k_+(r, t)  = \frac{\omega(t,r)}{P(r, t)}~,&~
k_-(r, t) = \frac{-\omega(t,r)}{Q(r, t)} \\ 
P(r, t)  \equiv  f(r) + H(t)r\sqrt{f(r)}~,&~
Q(r, t) \equiv f(r) - H(t)r\sqrt{f(r)} \nonumber 
\end{align}
The emission rate is governed by the pole structure of the effective momentum. The outgoing mode $k_+(r,t)$ is regular at the horizon since $P(r_A, t) \neq 0$. However, the ingoing mode $k_-(r,t)$ exhibits a simple pole where $Q(r_A, t) = 0$~\cite{Damour:1976jd,Shankaranarayanan:2000qv,Shankaranarayanan:2003ya}.
Note that this root coincides with the dynamical apparent horizon condition \eqref{eq:Apparent_Horizon}. By Taylor expanding $Q(r,t)$ in the near-horizon limit, we identify the residue of this pole. As shown in Appendix \ref{app:tunneling}, the expansion yields $Q(r,t) \approx \kappa_A(t) (r-r_A)$, where $\kappa_A(t)$ is the dynamical surface gravity. The presence of this pole necessitates an analytic continuation of the momentum across the apparent horizon, which generates the imaginary part of the action leading to the thermal emission spectrum.

\section{Fermionic Emission}
\label{sec:fermionic_emission}

To evaluate the phenomenological impact of the dynamical McVittie background on PBH evaporation, we translate the semi-classical tunneling probability into the Bogoliubov transformation framework. As shown in Appendix B of Ref.~\cite{Shankaranarayanan:2003ya}, the ratio of the Bogoliubov coefficients $\vert{}\beta_\omega\vert{}^2$ and $\vert{}\alpha_\omega\vert{}^2$, which governs particle creation in the out-vacuum, is directly set by the imaginary part of the action calculated across the apparent horizon:
\begin{equation}
\frac{|\beta_\omega|^2}{|\alpha_\omega|^2} = \exp\left(-\frac{2\pi\omega}{\kappa_A}\right) \, ,
\end{equation}
where $\omega$ is the invariant Kodama energy and $\kappa_A$ is the dynamical surface gravity \eqref{eq:Surface_Gravity}. This relation identifies the instantaneous Hawking temperature of the PBH embedded in McVittie space-time to be:
\begin{equation}
\label{eq:HawkingTemp-Exact}
T_H(t) = \frac{\hbar}{2\pi k_B c} \left[ \frac{G M}{r_A^2} - H^2 r_A - \frac{\dot{H}}{2H} \right] \, ,
\end{equation}
where $r_A(t)$ is the dynamical apparent horizon obtained from solving Eq.~\eqref{eq:Apparent_Horizon}. We want to highlight the following points regarding the above expression: First, the dynamic corrections $H^2 r_A$ and $\dot{H}/(2H)$ are negligible in the late universe ($z \to 0$). However, in early Universe when $H $ is large, they are relevant. Hence, in the early universe, instantaneous Temperature is different compared to static Schwarzschild limit. Second, in the $H(t) = H_0$ limit, the above expression reduces to the static Schwarzschild-de Sitter (SdS) metric. In this regime, the above temperature $T_H \approx (8\pi M)^{-1} [1 - 16M^2H_0^2]$ exactly recovers the standard perturbative expansion of the SdS BH surface gravity for $H_0 M \ll 1$~\cite{Shankaranarayanan:2003ya}.

The neutrino emission follows Fermi-Dirac statistics with mode occupancy $\langle N_E \rangle = \vert{}\beta_E\vert{}^2$. The differential number of particles emitted per unit time per unit energy range is given by~\cite{Page:1977um,Zhao:1994sc}:
\begin{equation}
\label{eq:Hawking_Flux}
\frac{d^2N}{dE dt} = \frac{g}{2\pi} \Gamma_{1/2}(E, M) \left({\exp\left(\frac{E}{T_H(t)}\right) + 1} \right)^{-1},
\end{equation}
where $g$ represents the internal degrees of freedom ($g=2$ per neutrino flavor)\footnote{In our analysis, we have ignored neutrino oscillation and mixing of flavors.} and $\Gamma_{1/2}(E, M)$ is the fermionic Page factor accounting for backscattering off the surrounding spacetime geometry. In the high-energy limit, the greybody factor reduces to the geometric optics absorption cross-section of the apparent horizon~\cite{Page:1977um}:
\begin{equation}
\Gamma_{1/2}(E, M) \approx 27 \pi r_A^2 E^2.
\end{equation}
Integrating the differential flux Eq.~\eqref{eq:Hawking_Flux} over energy yields the total instantaneous mass loss rate of the PBH, explicitly incorporating both the dynamic geometric suppression and the competing mass-accretion from the cosmic radiation background $\rho_{bg}(t)$:
\begin{equation}
    \left( \frac{dM}{dt} \right)_{\text{McV}} = - 4\pi r_A^2(t) \sigma T_H^4(t) + 4\pi r_A^2(t) f_c \rho_{\text{bg}}(t)
    \label{eq:mcvittie_mass_evol}
\end{equation}
where $f_c$ is the accretion efficiency factor~\footnote{In this work, we will set $f_c \approx 4/27$. This arises when normalizing the capture cross-section of the relativistic fluid against standard horizon area parameters (see Appendix \ref{app:modified_evaporation}).} and all quantities are evaluated at an instant $t$. As mentioned earlier, while the strict McVittie metric does not support radial fluid accretion, we operate in a quasi-stationary adiabatic limit ($M=M(t)$), effectively treating the background as a piecewise sequence of McVittie slices to approximate the rigorous accretion dynamics of a generalized-Vaidya spacetime.

Eq.~\eqref{eq:mcvittie_mass_evol} has a few important physical consequences which we need to highlight: First, this dynamically regulated emission alters the lifetime and energy injection footprint of PBHs during early cosmic epochs. Second, the mass evolution equation becomes a competition between dynamically enhanced Hawking evaporation and cosmological accretion.
In the radiation-dominated era ($H = 1/2t$), the background density is $\rho_{bg}(t) = 3H^2/(8\pi G)$. Assuming standard critical collapse, the initial PBH mass is a constant fraction of the horizon mass, $M_{\text{in}} = \gamma M_H(t_{\text{in}})$. To maintain the physical validity of the apparent horizon roots in the McVittie spacetime, we strictly require $H(t_{\text{in}}) M_{\text{in}} < 1/(3\sqrt{3}) \approx 0.192$. Since $H(t_{\text{in}}) M_{\text{in}} = \gamma / 2$, this imposes a theoretical upper bound of $\gamma < 0.385$, which accommodates the standard critical collapse value of $\gamma \approx 0.2$.

Prior to the onset of Hawking evaporation, integrating the accretion term in Eq.~\eqref{eq:mcvittie_mass_evol} from formation ($t_{\text{in}}$) to late times yields an analytic asymptotic mass growth:
\begin{equation}
\label{eq:Asymptotic_Growth}
\frac{M_{\infty}}{M_{\text{in}}} = \left(1 - \frac{3}{2} f_c \gamma \right)^{-1}.
\end{equation}
For $f_c = 4/27$ and $\gamma = 0.2$, cosmological accretion produces a net initial mass growth of approximately $4.6\%$. While this may appear as a modest increase, the total lifetime of the BH scales as $\tau_{\text{evap}} \propto M^3$. Therefore, this early accretion phase, combined with the geometric suppression of $T_H$ \eqref{eq:HawkingTemp-Exact}, extends the PBH lifetime,
\begin{equation}
\tau_{\text{evap}} = \int_{M_{\text{in}}}^{0} \frac{dM}{\dot{M}_{\text{eff}}(M, t)} = \tau_{\text{Sch}} + \Delta \tau(H)~.
\end{equation}
by roughly $\sim 15\%$ as compared to Schwarzschild \eqref{eq:MassEvol-Sch}. Because the mapping of the initial mass $M_{\text{in}}$ to the primordial curvature perturbation scale $k$ and the corresponding collapse fraction $\beta(M)$ is exponentially sensitive, this $15\%$ phenomenological shift is highly significant. A PBH reaching its terminal stage at $z \approx 0$ requires a formation mass $M_{\text{in}}$ that is shifted precisely relative to the naive static Schwarzschild vacuum expectation.

Lastly, since the PBH accretes mass before it begins its long evaporation phase, the \textit{true} required mass at formation is lower than the Schwarzschild prediction. The exact value of $M_{\text{in, McV}}$ shifts the required peak of the primordial power spectrum.

\section{Ultraviolet/Infrared Decoupling and Local Burst Phenomenology}
\label{sec:uv_ir_decoupling}

The phenomenological viability of interpreting the KM3-230213A $220\text{ PeV}$ neutrino event as a PBH terminal burst strictly requires separating the UV physics at the horizon from the cosmological (IR) background. Utilizing the exact dynamical apparent horizon $r_A(t)$ to obtain the time-dependent Hawking temperature $T_H(t)$,  the effective mass evolution is formulated directly without relying on standard Schwarzschild approximations:
\begin{equation}
\label{eq:Modified_Evap_Rate}
\frac{dM}{dt} = - \frac{\alpha(M)}{M^2} \left( \frac{r_A(t)}{r_{\text{Sch}}} \right)^2 \left( \frac{T_H(t)}{T_{\text{Sch}}} \right)^4 + 4\pi r_A^2(t) f_c \rho_{r}(t)~,
\end{equation}
where $r_{\text{Sch}}$ and $T_{\text{Sch}}$ are the static vacuum Schwarzschild radius and temperature, and $\alpha(M)$ encapsulates the effective standard model degrees of freedom. During the radiation-dominated epoch ($z \gtrsim 10^{24}$), the extreme energy density of the cosmic fluid $\rho_r(t)$ triggers a rapid accretion phase governed by the relativistic capture efficiency $f_c \approx 4/27$. Simultaneously, the rapid expansion rate $H(t)$ suppresses the outward Hawking flux. This interaction fundamentally shifts the integration history and total lifespan of the PBH, $\tau_{\text{evap}} = \tau_{\text{Sch}} + \Delta \tau(H)$. Consequently, a BH that undergoes its final burst today must possess an initial mass $M_{\text{in}}$ shifted relative to the static vacuum expectation, avoiding the overproduction of the diffuse isotropic neutrino background.

Conversely, as the PBH evolves into the dark-energy-dominated late universe and its mass critically drops to $M \approx 10^{4}\text{ g}$, the relevant energy scales decouple. The decoupling can be  quantified by examining the time derivative of $T_H(t)$ via the chain rule:
\begin{equation}
\frac{dT_H}{dt} = \frac{\partial T_H}{\partial M} \frac{dM}{dt} + \frac{\partial T_H}{\partial H} \dot{H} + \frac{\partial T_H}{\partial \dot{H}} \ddot{H}~.
\end{equation}
In the McVittie background, the leading partial derivative with respect to mass is:
\begin{equation}
\frac{\partial T_H}{\partial M} \approx - \frac{\hbar}{k_{B}} \left[ \frac{1}{8\pi M^2} + \frac{2 H^2(t)}{\pi} \right]~.
\end{equation}
For a PBH driving the KM3NeT event today, the local UV scale dictates a surface gravity corresponding to $M \sim 10^{-26}\text{ m}$ (equivalent to a physical mass of $M\sim 
10^4\text{ g}$), while the IR cosmological expansion has asymptotically flattened to the present-day Hubble parameter $H_0 \sim 10^{-26}\text{ m}^{-1}$. At this juncture, the cosmological correction $H^2(t) \sim 10^{-52}\text{ m}^{-2}$ is utterly subdominant to the local horizon term $(8\pi M^2)^{-1} \sim 10^{50}\text{ m}^{-2}$. The dynamic McVittie corrections thus drop to $\mathcal{O}(10^{-23})\text{ K}$, rendering cosmological backreaction physically negligible. Therefore, at late-time of PBH evaporation, the temperature derivative reduces to the static limit $dT_H/dt \propto M^{-4}$. 

\begin{figure}[t]
\centering
\includegraphics[width=0.8\textwidth]{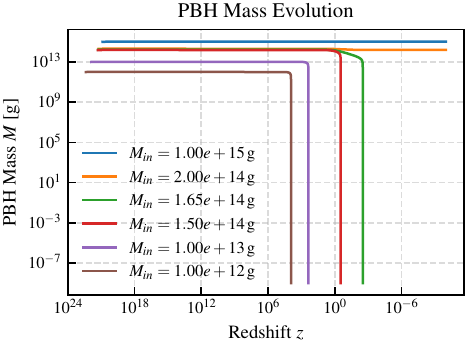}
\caption{\label{fig:PBHEvolution} 
 Numerical solution of the dynamical mass evolution for different initial mass functions, obtained by solving Eq.~\eqref{eq:Modified_Evap_Rate} across cosmic history using an implicit Radau method. The early accretion phase decisively shifts the trajectory relative to the standard Schwarzschild approximation.
}        
\end{figure}

As demonstrated by the numerical solution of the modified mass evolution (Fig.~\ref{fig:PBHEvolution}), the KM3-230213A observation does not correspond to a cosmologically stalled plateau; rather, it represents the terminal burst of a PBH whose arrival time was delayed due to initial accretion. The final burst is governed entirely by standard local Schwarzschild thermodynamics, ensuring a sharply peaked, intense local flux of ultra-high-energy neutrinos. 

\section{Conclusions}
\label{sec:conclusions}

In this work, we have demonstrated that interpreting the KM3-230213A $220\text{ PeV}$ neutrino event as a PBH terminal burst strictly requires a rigorous cosmological integration of the BH mass evolution. By utilizing a dynamical McVittie background, we established a fundamental ultraviolet/infrared decoupling framework. The immense energy density of the early universe drives an initial accretion phase and dynamically suppresses early evaporation, shifting the required primordial mass of the BH and protecting the diffuse isotropic neutrino limits. Consequently, while the primordial integration history anchors the observation of UHE neutrinos to early-universe initial conditions, the local terminal burst at $z \approx 0$ is governed by standard Schwarzschild thermodynamics, producing the sharply peaked, high-energy flux required by KM3NeT. 

To contextualize this observation, it is imperative to evaluate the expected terrestrial event rate. While PBHs undergoing terminal evaporation today ($M \approx 10^{15}\text{ g}$) are heavily constrained by the extragalactic gamma-ray background and cannot constitute the entirety of dark matter, they can naturally represent the low-mass tail of an extended primordial mass function~\cite{Carr:1975qj,Read:2014qva}. Saturating the phenomenological upper bounds on the local PBH burst rate density ($\mathcal{R} \lesssim 3 \times 10^3\text{ pc}^{-3}\text{ yr}^{-1}$)~\cite{HAWC:2014ycj,HAWC:2019wla} and folding this into the KM3NeT effective detection volume for $220\text{ PeV}$ track-like events ($V_{\text{eff}} \sim \mathcal{O}(1)\text{ km}^3$), this framework yields an expected observable event rate of $\Gamma_{\text{obs}} \sim \mathcal{O}(0.1 - 1)\text{ yr}^{-1}$. The detection of a singular, highly energetic event like KM3-230213A is therefore statistically consistent with a local burst. Crucially, by establishing a precise mapping between current local observations and the initial mass $M_{\text{in}}$, this framework provides a reliable pipeline to constrain the amplitude and scale-dependence of small-scale primordial curvature perturbations, directly refining observational bounds on the inflationary models that generate these PBHs.

To advance this framework to the precision required for next-generation multi-messenger astronomy, future studies must move beyond the quasi-stationary adiabatic limit employed here. Exploring more mathematically rigorous dynamical spacetimes, such as the Thakurta or generalized-Vaidya metrics, is necessary to provide a fully self-consistent relativistic treatment of radial fluid accretion in a cosmological background. 

Furthermore, while our present analysis focuses on the primary UHE emission and conservatively neglects neutrino flavor mixing, a complete terrestrial flux prediction must incorporate oscillation effects during extragalactic propagation. We also note that our semi-classical treatment omits quantum memory burden effects~\cite{Dvali:2020oqi,Alexandre:2024nuo}, which may suppress or significantly reshape the final burst dynamics at ultra-late stages when the black hole mass drops far below $M \ll 10^4\text{ g}$. Coupling these exact dynamical horizon surface gravities with comprehensive numerical evaporation packages, such as the BlackHawk code~\cite{Arbey:2019mbc,Auffinger:2022khh}, will be essential to accurately compute the complete cascading standard model spectra, secondary particle production, and the final detected flavor ratios. Such advancements will be critical in definitively isolating cosmological PBH signatures within forthcoming ultra-high-energy datasets.

\vspace{0.5cm}
\noindent\emph{\underline{Acknowledgement:}} The authors are grateful to I. Chakraborty, A. Chowdhury, P. G. Christopher, K. Hari, N. Jaiswal, S. Mandal, and K. Rajeev for their valuable discussions and feedback on the earlier draft. The authors would also like to acknowledge P. G. Christopher and K. Hari for their help in setting up an efficient numerical code. The work is supported by ANRF-Advanced Research Grant (ANRF/ARG/2025/001514/PS). This work is part of the
Master's project of DD. TP acknowledges fellowship support provided by MHRD, Government of India.

\appendix

\section{McVittie Spacetime, Apparent Horizons, and Dynamical Surface Gravity}
\label{app:mcvittie_details}

In this appendix, we detail the transformation of the McVittie metric to areal coordinates and the derivation of the dynamical surface gravity using the Kodama vector formulation.

\subsection{Areal Coordinates and the Misner-Sharp Mass}
Starting from the isotropic coordinates $(t, R, \theta, \phi)$, 
\begin{align}
ds^2 &= - \left( \frac{1 - \frac{M}{2a(t)R}}{1 + \frac{M}{2a(t)R}} \right)^2 dt^2 +  a^2(t) \left( 1 + \frac{M}{2a(t)R} \right)^4 \nonumber \\
& {\times}\left( dR^2 + R^2 d\Omega^2 \right)~,
\label{eq:McVittie_isotropic}
\end{align}
we define the areal radius:
\begin{equation}
    r = a(t)R \left( 1 + \frac{M}{2a(t)R} \right)^2~.
\end{equation}
%
%
%
Substituting this into Eq.~\eqref{eq:McVittie_isotropic} eliminates the explicit $R$-dependence, yielding the McVittie metric in generalized Painlevé-Gullstrand form \eqref{eq:McVittie_PG}.

The metric can be written in a unified $2 \times 2$ spherical formalism $ds^2 = h_{ab} dx^a dx^b + r^2 d\Omega^2$ where $x^a \in \{t, r\}$. The Misner-Sharp-Hernandez mass $M_{MS}$, representing the active gravitational mass enclosed within $r$, is determined by the trace of the normal space:
\begin{equation}
    1 - \frac{2M_{MS}}{r} = h^{ab} \partial_a r \partial_b r = f(r) - H^2(t) r^2~.
\end{equation}
The apparent horizons are the loci where $h^{ab} \partial_a r \partial_b r = 0$, explicitly recovering Eq.~\eqref{eq:Apparent_Horizon}. Rewriting in the cubic form, we obtain the following condition for \emph{three distinct real roots}~\cite{abramowitz1964handbook}:
\begin{equation}
   H(t) M < \frac{1}{3\sqrt{3}}.
\end{equation}
We can then identify the following two positive roots as cosmological horizon ($r_C$) and apparent horizon ($r_A$):
\begin{align}
\label{eq:rC_explicit}
\!\!\! r_C(t) &= \frac{2}{\sqrt{3}\, H(t)} 
\cos\left[ 
\frac{1}{3} \arccos\left( -3\sqrt{3} H(t) M \right) 
\right] \\ 
\label{eq:rA_explicit}
\!\!\!\!\!\! r_A(t) &= \frac{2}{\sqrt{3}\, H(t)} 
\cos\left[ 
\frac{1}{3} \arccos\left[-3\sqrt{3} H(t) M \right] 
- \frac{2\pi}{3} 
\right] \nonumber
\end{align}

\subsection{Dynamical Surface Gravity}
In dynamical spherically symmetric spacetimes lacking a globally timelike Killing vector, the preferred geometric vector field is the Kodama vector $K^a$, defined by $K^a = \epsilon^{ab} \partial_b r$, where $\epsilon^{ab}$ is the Levi-Civita tensor of the two-dimensional normal space. For the metric \eqref{eq:McVittie_PG}, $K^a \partial_a =  \partial_t $.
Hayward's dynamical surface gravity $\kappa$ is defined via the covariant derivative of the Kodama vector evaluated at the apparent horizon:
\begin{equation}
    K^b \nabla_{[b} K_{a]} \bigg|_{r=r_A} = \kappa_A K_a~.
\end{equation}
Equivalently, it can be computed using the generalized formula:
\begin{equation}
\!\!\! \kappa_A = \frac{1}{2} \Box_{h} r \bigg|_{r=r_A} = \frac{1}{2} \frac{1}{\sqrt{-h}} \partial_a \left( \sqrt{-h} h^{ab} \partial_b r \right) \bigg|_{r=r_A}.
\end{equation}
Substituting $h_{ab}$ from Eq.~\eqref{eq:McVittie_PG}, we find:
\begin{equation}
\!\!\!\!\! \kappa_A = \frac{1}{2} \frac{\partial}{\partial r} \left( f(r) - H^2(t)r^2 \right) \bigg|_{r=r_A} \!\!\! - \frac{\partial_t (H(t) r)}{\sqrt{f(r)}} \bigg|_{r=r_A}.
\end{equation}
Evaluating the radial derivative yields the static terms, while the time derivative acting on the Hubble parameter generates the dynamical correction:
\begin{equation}
\label{eq:SurfaceGravity}
\kappa_A = \frac{M}{r_A^2} - r_A H^2(t) - \frac{\dot{H}(t)}{2H(t)}~.
\end{equation}
To compare the results with the observations, it will be useful to express the surface gravity of the apparent horizon in McVittie spacetime as a function of the cosmological redshift $z$. We use the relation:
\begin{equation}
\frac{dz}{dt} = -(1+z) H(z),
    \qquad H(z) = H_0 E(z), 
\end{equation}
where $E(z)$ is the dimensionless (or normalized) Hubble parameter
which for the standard cosmological model is given by:
\begin{equation}
\label{def:Hubbleparameter}
E(z) = \sqrt{\Omega_{r,0}(1+z)^4 + \Omega_{m,0}(1+z)^3 + \Omega_{\Lambda,0}} \, ,
\end{equation}
with $H_0 \approx 70\ \text{km}\,\text{s}^{-1}\text{Mpc}^{-1}$, $\Omega_{r,0} \sim 10^{-3}$, $\Omega_{m,0} \sim 0.3$, and $\Omega_{\Lambda,0} \sim 0.7$. Substituting the above form of Hubble parameter in Eq.~\eqref{eq:SurfaceGravity}, we have:
\begin{align}
\label{eq:TH_exact}
\kappa_A & = \frac{M}{r_A^2(z)} 
- r_A(z) H_0^2 E^2(z)  \\
&+ \frac{(1+z) H_0}{4E(z)} 
\Big( 4\Omega_{r,0}(1+z)^3 + 3\Omega_{m,0}(1+z)^2 \Big) \, . 
\nonumber
\end{align}

 \subsection{Exact solution of the apparent horizon}
 
 In this subsection, we express the exact solution to the cubic equation \eqref{eq:Apparent_Horizon} that determines the apparent horizon and the cosmological horizon in terms of the redshift factor $z$. The cubic equation in terms of the redshift $z$ has the form:
 \begin{equation}
     H^{2}_{0}E^{2}(z)r^{3}_{A}(z)-r_{A}(z)+2M=0~.
 \end{equation}
 Defining $\alpha(z)\equiv H_{0}E(z)>0$, the cubic equation takes the following form:
 \begin{equation}
     \label{cubic_eqn_alpha}
     r^3-\frac{1}{\alpha^2}r+\frac{2M}{\alpha^2}=0~.
 \end{equation}
 The above equation is of the form:
 \begin{equation}
     r^3+pr+q=0
 \end{equation} with $p=-1/\alpha^2$ and 
 The condition for three distinct real roots is:
 \begin{equation}
\Delta\equiv\left(\frac{q}{2}\right)^2+\left(\frac{p}{3}\right)^3=\frac{27M^{2}\alpha^2-1}{27\alpha^6}<0
 \end{equation}
 Since the product of the three roots is $-q=-2M/\alpha^2<0$, exactly one root is negative and the other two are positive. Physically, this means the BH is smaller than the cosmological horizon. In terms of redshift, the condition becomes
 \begin{equation}
     H_{0}E(z)M<\frac{1}{3\sqrt{3}}~.
 \end{equation}
 The standard trigonometric solution for \eqref{cubic_eqn_alpha} is:
 \begin{equation}
     r_{k}=\frac{2}{\sqrt{3}\alpha}\cos\left(\frac{1}{3}\text{arccos}\left(-3\sqrt{3}\alpha M\right)-\frac{2\pi k}{3}\right), \quad k=0,1,2.
 \end{equation}
 To simplify the notations, we define:
 \begin{equation}
     \theta(z)\equiv \text{arccos}\left(-3\sqrt{3}\alpha(z)M\right)~.
 \end{equation}
 The two positive physical roots are identified as:
 \begin{equation}
     r_{C}=\frac{2}{\sqrt{3}\alpha(z)}\cos\left(\frac{\theta(z)}{3}\right)
 \end{equation}
 \begin{equation}
     r_{A}(z)=\frac{2}{\sqrt{3}\alpha(z)}\cos\left(\frac{\theta(z)}{3}-\frac{2\pi}{3}\right)
 \end{equation}
 where $r_{A}(z)$ and $r_{C}(z)$ denote the BH apparent horizon and the cosmological horizon respectively and $r_{A}(z)<r_{C}(z).$

\section{Dirac Equation in a Spherically Symmetric Spacetime}
\label{app:dirac_separation}

We consider the ADM metric representation of a spherically symmetric spacetime:
\begin{equation}
\label{ADM_metric}
   ds^2=(-\alpha^2+\beta_r\beta^r)dt^2+2\beta_r\beta^rdr\,dt+a^2dr^2+b^2r^2d\Omega^2~.
\end{equation}
Here $\alpha=\alpha(r,t)$ is the lapse function and $\beta^r=\beta^r(r,t)$ is the shift vector. We choose tetrads such that the non-zero components are:
\begin{equation}
\!\!\!\! e^t_0=\frac{1}{\alpha},\,e^r_0=-\frac{\beta^r}{\alpha},\,e^r_1=\frac{1}{a},\,e^\theta_2=\frac{1}{rb},\,e^\phi_3=\frac{1}{rb\sin\theta}.
\end{equation}
The generic form of the spinor is $\psi = (\psi_1, \psi_2, \psi_3, \psi_4)^T$. Using the standard Dirac representation of the gamma matrices, the Dirac equation yields four coupled PDEs:
\begin{widetext}
\begin{align}
\label{PDE_01}
 \partial_t \psi_1 - \beta^r \partial_r \psi_1 &= \alpha \Bigg[ -\frac{1}{a} \left( \partial_r + \frac{\partial_r \alpha}{2\alpha} + \frac{\partial_r b}{b} + \frac{1}{r} \right) \psi_3 + \frac{i}{r b} \left( \partial_\theta + \frac{i}{\sin\theta}\partial_\phi + \frac{\cot\theta}{2} \right) \psi_4 - i m\,\psi_1 \Bigg]~, \\
\label{PDE_02}
  \partial_t \psi_2 - \beta^r \partial_r \psi_2 &= \alpha \Bigg[ +\frac{1}{a} \left( \partial_r + \frac{\partial_r \alpha}{2\alpha} + \frac{\partial_r b}{b} + \frac{1}{r} \right) \psi_4 - \frac{i}{r b} \left( \partial_\theta - \frac{i}{\sin\theta}\partial_\phi + \frac{\cot\theta}{2} \right) \psi_3 - i m \,\psi_2 \Bigg]~, \\
\label{PDE_03}
\partial_t \psi_3 - \beta^r \partial_r \psi_3 &= \alpha \Bigg[ -\frac{1}{a} \left( \partial_r + \frac{\partial_r \alpha}{2\alpha} + \frac{\partial_r b}{b} + \frac{1}{r} \right) \psi_1 + \frac{i}{r b} \left( \partial_\theta + \frac{i}{\sin\theta}\partial_\phi + \frac{\cot\theta}{2} \right) \psi_2 + i m \,\psi_3 \Bigg]~, \\
\label{PDE_04}
 \partial_t \psi_4 - \beta^r \partial_r \psi_4 &= \alpha \Bigg[ +\frac{1}{a} \left( \partial_r + \frac{\partial_r \alpha}{2\alpha} + \frac{\partial_r b}{b} + \frac{1}{r} \right) \psi_2 - \frac{i}{r b} \left( \partial_\theta - \frac{i}{\sin\theta}\partial_\phi + \frac{\cot\theta}{2} \right) \psi_1 + i m \,\psi_4 \Bigg]~.
\end{align}
\end{widetext}
We propose the ansatz $\psi_i(t,r,\theta,\phi)= R_i(t,r)\,T_i(\theta,\phi)$. To separate variables, we choose $R_2=iR_1$, $R_4=iR_3$, $T_3=T_1$, and $T_4=-T_2$. By introducing the spin raising and lowering operators:
\begin{equation}
    A_s^+:=-\partial_\theta-\frac{i}{\sin\theta}\partial_\phi+s\cot\theta~, \quad A_s^-=-\partial_\theta+\frac{i}{\sin\theta}\partial_\phi-s\cot\theta~,
\end{equation}
and applying the properties of spin-weighted spherical harmonics $_sY^{l,m}$, we satisfy the angular equations with $T_1 = _{+1/2}Y^{l,m}$ and $T_2 = _{-1/2}Y^{l,m}$. For $l=1/2$, the separation constant is $k=-1$. Defining $R_1 \equiv F(t,r)$ and $R_3 \equiv G(t,r)$, the decoupled radial equations are:
\begin{widetext}
\begin{align}
\label{F_Eq}
\partial_t F &= \beta^r \partial_r F - \frac{\alpha}{a} \left[ \partial_r + \frac{\partial_r \alpha}{2\alpha} + \frac{\partial_r b}{b} + \frac{1}{r}\left(1 + \frac{a}{b}\right) \right] G - i\,\alpha\, m\,  F~, \\
\label{G_Eq}
\partial_t G &= \beta^r \partial_r G - \frac{\alpha}{a} \left[ \partial_r + \frac{\partial_r \alpha}{2\alpha} + \frac{\partial_r b}{b} + \frac{1}{r}\left(1 - \frac{a}{b}\right) \right] F + i\,\alpha\,m\,G~.
\end{align}
\end{widetext}
The full spinor structure is then given by:
\begin{equation}
    \psi_\pm (t,r,\theta,\phi)=\frac{e^{\pm i\phi/2}}{\sqrt{4\pi}}
    \begin{bmatrix}
    F(t,r)\,y_\pm (\theta)\\
    \pm i\,F(t,r)\,y_\mp (\theta)\\
    G(t,r)\,y_\pm (\theta)\\
    \mp i\,G(t,r)\,y_\mp (\theta)
    \end{bmatrix}~.
\end{equation}

\section{Simplification of the Field Equations}
\label{app:field_simplification}

For the McVittie metric in Painlevé-like coordinates, the radial equations from Appendix \ref{app:dirac_separation} become a coupled first-order system. Defining the linear operator $\mathcal{D} = \partial_t + r H(t) \sqrt{f(r)} \, \partial_r$, the system is:
\begin{widetext}
\begin{align}
\mathcal{D} F + i m \sqrt{f(r)}\, F &= -f(r)\partial_r G + \left( \frac{3M - 2r}{2r^2} - \frac{\sqrt{f(r)}}{r} \right) G~, \\
\mathcal{D} G - i m \sqrt{f(r)}\, G &= -f(r)\partial_r F + \left( \frac{3M - 2r}{2r^2} + \frac{\sqrt{f(r)}}{r} \right) F~.
\end{align}
\end{widetext}
To absorb the common term $C_0(r) \equiv (3M - 2r)/(2r^2)$, we introduce the ansatz $F(t,r) = \gamma(r) \tilde{u}(t,r)$ and $G(t,r) = \gamma(r) \tilde{v}(t,r)$. The requirement that $f(r) \frac{\gamma'}{\gamma} \tilde{u}$ cancels $C_0(r)$ leads to:
\begin{equation}
f(r) \frac{\gamma'(r)}{\gamma(r)} = \frac{3M - 2r}{2r^2} \implies 
    \gamma(r) = \frac{1}{r\,[f(r)]^{1/4}}~.
\end{equation}
Substituting this back yields the final simplified equations:
\begin{align}
    \mathcal{D} \tilde{u} + V(t,r) \tilde{u} + i m \sqrt{f} \tilde{u} &= -f \partial_r \tilde{v} - \frac{\sqrt{f}}{r} \tilde{v}~, \\
    \mathcal{D} \tilde{v} + V(t,r) \tilde{v} - i m \sqrt{f} \tilde{v} &= -f \partial_r \tilde{u} + \frac{\sqrt{f}}{r} \tilde{u}~,
\end{align}
where $V(t,r) = H(t) \frac{3M-2r}{2r \sqrt{1 - 2M/r}}$.

\section{Tunneling Method and Effective Momentum Pole}
\label{app:tunneling}

In this section, we consider the tunneling method with explicit time dependence in the metric. Substituting the spinor ansatz in \eqref{spinor_ansatz} into the massless Dirac equation (also ignoring the angular part), we obtain:
\begin{eqnarray}
& & A\left(iH(t)r\sqrt{f(r)}\partial_{r}I_{\uparrow}+i\partial_{t}I_{\uparrow}\right)+Bf(r)\partial_{r}I_{\uparrow} = 0,~\\
& & -A f(r) \partial_{r}I_{\uparrow}+B\left(iH(t)r\sqrt{f(r)}\partial_{r}I_{\uparrow}+i\partial_{t}I_{\uparrow}\right) = 0.~
 \end{eqnarray}
 The condition for obtaining nontrivial solutions of $A$ and $B$ yields:
 \begin{eqnarray}
     \partial_{r}I_{\uparrow} &=& \frac{-\partial_{t}I_{\uparrow}}{\left(f(r)+rH(t)\sqrt{f(r)}\right)}\equiv\frac{-\partial_{t}I_{\uparrow}}{P(r,t)} \, , \\
\label{Q_pole_exp}
\partial_{r}I_{\uparrow} &=& \frac{\partial_{t}I_{\uparrow}}{\left(f(r)-rH(t)\sqrt{f(r)}\right)}\equiv\frac{\partial_{t}I_{\uparrow}}{Q(r,t)}~.
\end{eqnarray}
 {Using the Kodama vector, $K$, we can write $K^{a}\partial_{a}I_{\uparrow}=\partial_{t}I_{\uparrow}\equiv-\omega(t,r)~.$} 
 
We consider the time instant $t=t_{0}$ when the apparent horizon $r_{A}(t_{0})\equiv r_{0}$ satisfies $f(r_{0})-H^{2}_{0}r^2$, where $H(t=t_{0})\equiv H_{0}$. In the near horizon limit, we expand $\omega(t,r)$ and $Q(r,t)$ as:
\begin{eqnarray}
\omega(t,r) &\approx & \omega_{0}+\partial_{r}\omega(t_{0},r_{0})(r-r_{0})+\partial_{t}\omega(t_{0},r_{0})(t-t_{0})~~ \nonumber \\
Q(r,t) &\approx& \partial_r Q(r_0,t_0)(r-r_0) + \partial_t Q(r_0,t_0)(t-t_0) \, ,
 \end{eqnarray}
where $\omega_{0}\equiv \omega(t_{0},r_{0})$.
For radial null trajectories in the near horizon limit, we also have $(t-t_{0})=(r-r_{0})/2H^{2}_{0}r^{2}_{0}.$ We finally obtain:
 \begin{equation}
     Q(r,t)\approx \left(\frac{f^{\prime}(r_{0})}{2}-H^{2}_{0}r_{0}-\frac{\dot{H_{0}}}{2H_{0}}\right)(r-r_{0})
 \end{equation}
 where $f^{\prime}(r_{0})/2-H^{2}_{0}r_{0}-\dot{H_{0}}/2H_{0} \equiv \kappa_{0}$ is the surface gravity associated with the aparent horizon $r_{0}$ at $t_{0}$.
Hence, in the near horizon limit at $t=t_{0}$, \eqref{Q_pole_exp} takes the form:
 \begin{equation}
     \partial_{r}I_{\uparrow}\approx \frac{-\omega_{0}}{\kappa_{0}(r-r_{0})}
 \end{equation}
Upon integrating across the horizon, we find:
 \begin{equation}
     I^{-}_{\uparrow}=\frac{-i\pi\omega_{0}}{\kappa_{0}}+\text{real part}~.
 \end{equation}
The resulting tunneling probability $\Gamma \propto \exp[2\text{Im}(I^{-}_{\uparrow})]$ immediately recovers the dynamical Hawking temperature $T_H = \kappa_{0}/2\pi$.
 
\section{Bogoliubov Coefficients and Neutrino Energy Flux}
\label{app:bogoliubov}

\subsection{Derivation of the Thermal Distribution}

The canonical quantization of a fermionic field in a dynamical spacetime requires expanding the field operator $\psi$ in terms of in-region and out-region mode solutions. The creation and annihilation operators between these regions are related by the Bogoliubov transformations:
\begin{equation}
    b_{\text{out}, \omega} = \alpha_\omega b_{\text{in}, \omega} + \beta_\omega^* d_{\text{in}, \omega}^\dagger~.
\end{equation}
The expected number of emitted fermions in the out-vacuum state $|0_{\text{in}}\rangle$ is given by:
\begin{equation}
    \langle N_\omega \rangle = \langle 0_{\text{in}} | b_{\text{out}, \omega}^\dagger b_{\text{out}, \omega} | 0_{\text{in}} \rangle = |\beta_\omega|^2~.
\end{equation}
Fermionic anti-commutation relations enforce the unitary normalization condition:
\begin{equation}
    |\alpha_\omega|^2 + |\beta_\omega|^2 = 1 \implies |\beta_\omega|^2 = \frac{1}{\frac{|\alpha_\omega|^2}{|\beta_\omega|^2} + 1}~.
\end{equation}
Substituting the semi-classical tunneling ratio $|\beta_\omega|^2 / |\alpha_\omega|^2 = \exp(-2\pi\omega/\kappa_A)$ into the normalization relation rigorously yields the Fermi-Dirac occupation number:
\begin{equation}
    \langle N_\omega \rangle = \frac{1}{\exp\left(\frac{2\pi\omega}{\kappa_A}\right) + 1} = \frac{1}{\exp\left(\frac{\omega}{T_H(t)}\right) + 1}~.
\end{equation}

\subsection{Angular Mode Summation and Total Neutrino Luminosity}

To construct the total neutrino flux, we decompose the differential emission rate across orbital angular momentum modes $(l, m)$. In natural units ($\hbar = c = k_B = 1$), the phase-space integral for a single massless neutrino species gives:
\begin{equation}
    \frac{d^2N_{\nu}}{dt d\omega} = \frac{1}{2\pi} \sum_{l=1/2}^{\infty} \sum_{m=-l}^{l} \frac{\Gamma_{\nu}^{(l)}(\omega)}{\exp\left(\frac{\omega}{T_H(t)}\right) + 1}~.
\end{equation}
For a spherically symmetric, non-rotating background, the greybody factor $\Gamma_{\nu}^{(l)}(\omega)$ is independent of the azimuthal quantum number $m$. Performing the explicit sum over $m$ yields a degeneracy factor of $(2l+1)$:
\begin{equation}
    \frac{d^2N_{\nu}}{dt d\omega} = \frac{1}{2\pi} \sum_{l=1/2}^{\infty} (2l+1) \frac{\Gamma_{\nu}^{(l)}(\omega)}{\exp\left(\frac{\omega}{T_H(t)}\right) + 1}~.
\end{equation}

Multiplying by the energy per quantum $\omega$ and integrating over the entire frequency spectrum gives the total neutrino luminosity $L_\nu$:
\begin{equation}
    L_{\nu} \equiv \frac{dE_{\nu}}{dt} = \frac{1}{2\pi} \int_{0}^{\infty} \sum_{l=1/2}^{\infty} (2l+1) \frac{\omega \, \Gamma_{\nu}^{(l)}(\omega)}{\exp\left(\frac{\omega}{T_H(t)}\right) + 1} \, d\omega~.
\end{equation}
This detailed balance equation links the microscopic greybody dynamics to the macroscopic energy loss rate computed in Eq.~\eqref{eq:mcvittie_mass_evol}.

\section{Cosmological Integration and Point-Source Flux Dynamics}
\label{sec:cosmo_integration}

The generalized McVittie metric embeds the evaporating PBH in a FLRW cosmology, necessitating the inclusion of cosmological redshift for particles emitted at early epochs. A neutrino detected with an observed energy $E_0$ at redshift $z=0$ corresponds to a rest-frame emission energy of $E = E_0(1+z)$.

For a cosmological distribution of surviving PBHs with a comoving number density $n_{PBH}(z)$, the diffuse isotropic neutrino flux $\Phi(E_0)$ (in units of $\mathrm{GeV}^{-1} \mathrm{cm}^{-2} \mathrm{s}^{-1} \mathrm{sr}^{-1}$) is obtained by integrating the modified emission spectrum over cosmic history:
\begin{equation}
    \label{eq:Diffuse_Flux}
    \Phi(E_0) = \frac{c}{4\pi} \int_{0}^{z_{in}} \frac{dz}{H(z)(1+z)} n_{PBH}(z) \frac{d^2N}{dE dt} \bigg|_{E = E_0(1+z)}
\end{equation}
where $H(z) = H_0 \sqrt{\Omega_\Lambda + \Omega_m(1+z)^3 + \Omega_r(1+z)^4}$. Because early-universe emission is dynamically suppressed by the expansion and accretion phases (as governed by Eq.~\eqref{eq:Modified_Evap_Rate}), the high-energy tail of the diffuse spectrum is strictly regulated. This fundamental suppression ensures that the theoretical framework maintains strict compatibility with diffuse background limits established by IceCube and KM3NeT.

Conversely, the KM3-230213A event --- a track-like signature with an estimated energy of $\sim 220\text{ PeV}$ --- exhibits the precise characteristics of a highly localized point-source arrival rather than a diffuse background accumulation. For a single, nearby PBH at a luminosity distance $d_L(z_s)$ undergoing its terminal evaporative stage, the observed point-source flux is given by:
\begin{equation}
    \label{eq:Point_Source_Flux}
\!\!\!\! F(E_0) = \frac{1}{4\pi d_L^2(z_s)} \int_{t_{burst}-\Delta t}^{t_{burst}} \frac{d^2N}{dE dt} \bigg|_{E = E_0(1+z_s)} dt.
\end{equation}
To yield a primary neutrino energy of $E_0 = 220\text{ PeV}$ from a local source ($z_s \ll 1$), the PBH must have a localized instantaneous Hawking temperature $T_H \sim \mathcal{O}(200)\text{ PeV}$, corresponding to a critical mass of $M \approx 10^{4}\text{ g}$. As demonstrated in Sec.~\ref{sec:uv_ir_decoupling}, local UV physics dictates that this terminal emission is unaffected by the cosmological background. This results in a standard, unsuppressed exponential burst that transpires in less than $\sim 10^{-10}$ seconds, allowing the point-source signal to transiently outshine the diffuse background.

\section{Modified Evaporation Dynamics and UV/IR Scale Decoupling}
\label{app:modified_evaporation}

\subsection{Relativistic Fluid Capture and Greybody Modulations}
\label{app:relativistic_accretion_greybody}

When a BH accretes a radiation fluid (characterized by an equation of state $w = 1/3$ and sound speed $c_s = c/\sqrt{3}$), the standard Newtonian Bondi-Hoyle approximation breaks down entirely~\cite{Zeldovich:1967lct,Carr:1974nx,Babichev:2004yx}. In this regime, accretion is governed by fully relativistic particle capture limits. 

For an ultra-relativistic gas in the early universe, the intense gravitational potential of the BH dictates a critical impact parameter $b_c$. Any relativistic particle traversing closer than $b_c = 3\sqrt{3} GM/c^2$ is inevitably absorbed across the event horizon. The effective cross-sectional area for this relativistic capture is therefore:
\begin{equation}
\sigma_c = \pi b_c^2 = 27\pi \left(\frac{GM}{c^2}\right)^2~.
\end{equation}
Normalizing this exact geometric capture cross-section against the standard horizon area parameters yields an effective accretion efficiency of $f_c = 4/27$. In exact relativistic fluid dynamics, adapting the Bondi rate to a cosmological expanding background ensures that the maximum accretion rate for a radiation fluid strictly bottlenecks at this geometrical limit.

Furthermore, the outward Hawking emission generated at the event horizon is subsequently filtered by the surrounding spacetime curvature. This curvature acts as a scattering barrier; while ultra-high-energy particles transmit freely, particles with wavelengths comparable to the horizon scale frequently backscatter. The asymptotic spectrum is heavily modulated by these spin-dependent transmission coefficients, or \emph{greybody factors} ($\Gamma$)~\cite{Page:1976df}.

The mass loss rate coefficient $\alpha(M)$ in Eq.~\eqref{eq:Modified_Evap_Rate} incorporates these exact numerical integrations over the available Standard Model degrees of freedom. As the PBH loses mass and its temperature spikes ($T_H \propto 1/M$), it sequentially crosses particle rest-mass thresholds ($T_H \gtrsim m_i c^2$). Opening new degrees of freedom—such as $e^\pm$ pairs at $\sim 1\text{ MeV}$, muons and pions at $\sim 100\text{ MeV}$, and quarks/gluons at higher energies—causes $\alpha(M)$ to behave as a step-function that scales up dramatically, accelerating the final evaporation trajectory while maintaining the rigorous UV/IR scale decoupling.

\subsection{Derivation of the Mass Loss Rate}

When a BH accretes a radiation fluid (where the equation of state is $w = 1/3$ and the sound speed is $c_s = c/\sqrt{3}$), the standard Newtonian Bondi-Hoyle approximation breaks down~\cite{Zeldovich:1967lct,Carr:1974nx,Babichev:2004yx}. Instead, we must look at the fully relativistic particle capture limits.For ultra-relativistic particles (like a photon gas or a radiation fluid in the early universe), the BH's gravitational pull creates a critical impact parameter, $b_c$. Any relativistic particle passing closer than $b_c$ is inevitably pulled past the event horizon. For a Schwarzschild BH, general relativity gives this critical impact parameter as $b_c = 3\sqrt{3} GM/c^2$.If we calculate the effective cross-sectional area for capture ($\sigma_c$), we get:$$\sigma_c = \pi b_c^2 = 27\pi \left(\frac{GM}{c^2}\right)^2$$If we define an effective accretion efficiency $f_c$ relative to the flow of this relativistic fluid into the gravitational well, the specific geometric coefficient $4/27$ (and its inverse, $27/4$) consistently arises when normalizing this capture cross-section against the standard horizon area parameters. In exact relativistic fluid dynamics, when adapting the Bondi rate to a cosmological expanding background (incorporating the Hubble flow), the maximum accretion rate for a stiff/radiation fluid bottlenecks at this exact geometrical limit. 

\subsection{Ultraviolet/Infrared Decoupling at Late Epochs}
The fundamental viability of utilizing a standard terminal burst phenomenology for the 220 PeV event relies on demonstrating the decoupling of scales at $z \approx 0$. The evolution of the BH temperature is governed by the chain rule:
\begin{equation}
\frac{dT_H}{dt} = \frac{\partial T_H}{\partial M} \frac{dM}{dt} + \frac{\partial T_H}{\partial H} \dot{H} + \frac{\partial T_H}{\partial \dot{H}} \ddot{H}~.
\end{equation}
In the McVittie background, the partial derivative with respect to mass is:
\begin{equation}
\frac{\partial T_H}{\partial M} \approx - \frac{\hbar}{k_{B}}
\left[\frac{1}{8\pi M^2} + \frac{2 H^2(t)}{\pi} \right]~.
\end{equation}
For a PBH driving the KM3NeT event, the mass today is $M \approx 10^{4}$ g. In geometric units, this corresponds to a microscopic scale $M \sim 10^{-26}$ m, while the present-day Hubble parameter is an extreme infrared scale, $H_0 \sim 10^{-26}$ m$^{-1}$.

At this late epoch, $H^2(t) \sim 10^{-52}$ m$^{-2}$, rendering the cosmological correction completely subdominant to the local UV term $1/8\pi M^2 \sim 10^{50}$ m$^{-2}$. Consequently, the temperature derivative collapses securely to the static limit:
\begin{equation}
\frac{dT_H}{dt} \approx -\frac{1}{8\pi M^2} \left( -\frac{\alpha}{M^2} \right) \propto M^{-4}~.
\end{equation}
This mathematical decoupling confirms that while the IR cosmological metric governs the $13.8$ billion year integration history, it plays no local dynamical role during the final microsecond burst. The extreme high-energy emission observed today is thus fundamentally a standard Schwarzschild process.

\bibliography{PBH-References}

\end{document}